\documentclass[11pt]{llncs}

\usepackage[english]{babel}
\usepackage[dvips]{graphicx}
\usepackage{epstopdf,enumerate,amsmath,amssymb,color,graphicx,bbm,float}

\begin{document}
\title{Long distance quantum key distribution with continuous variables}
\author{Anthony Leverrier\inst{1,2} \and 
        Philippe Grangier\inst{3}}
\institute{ICFO-Institut de Cienc\`es Fot\`oniques,\\ Mediterranean Technology Park, 08860 Castelldefels (Barcelona), Spain
\and Institut Telecom / Telecom ParisTech, CNRS LTCI,\\
  46, rue Barrault, 75634 Paris Cedex 13, France
   \and
Laboratoire Charles Fabry, Institut d'Optique, CNRS, Univ. Paris-Sud,\\
  Campus Polytechnique, RD 128, 91127 Palaiseau Cedex, France}

\maketitle    
   
\begin{abstract}
We present a continuous-variable quantum key distribution protocol combining a continuous but slightly non-Gaussian modulation together with a efficient reverse reconciliation scheme. We establish the security of this protocol against collective attacks which correspond to a linear quantum channel. In particular, all Gaussian attacks are considered in our framework. We show that this protocol outperforms all known practical protocols, even taking into account finite size effects.
\end{abstract}

\section{Introduction}

Quantum key distribution (QKD) is a cryptographic primitive allowing two distant parties, Alice and Bob, to establish a secret key in an untrusted environment controlled by some eavesdropper, Eve \cite{SBC08}. One of the great interests of QKD is that it can be implemented with present day technology, at least for reasonable distances.

Whereas discrete-variable protocols, such as BB84 \cite{bb84}, are quite resistant to losses (experiments over more than 200 km have been achieved \cite{SWV09}), continuous-variable protocols do not seem to display the same quality: the present experimental record is around 25 km \cite{LBG07,XZV09}, although recent theoretical results suggest than 100 km should be achievable \cite{LG09,LG10}. On the other hand, continuous-variable (CV) QKD does not require specific equipment such as single-photon counters and can be implemented with off-the-shelf telecom components. For this reason, it is of great importance to find the protocols with the highest resistance to losses. In this paper we introduce such a protocol.

Before describing the new protocol, let us first recall the main ideas of CVQKD, and explain the origin of its sensitivity to losses.
The basic idea is to encode information on continuous variables in phase space to perform QKD \cite{ral99}. This can be achieved with coherent states: the information is simply encoded in their displacement vectors and can be recovered by Bob thanks to homodyne or heterodyne detection (homodyne corresponds to the case where one random quadrature is measured, heterodyne means that both quadratures are measured). Two main categories of modulation have been considered in the literature: a continuous Gaussian modulation, which maximizes the mutual information $I_{AB}$ between Alice and Bob, and discrete modulations (mainly consisting of either 2 or 4 states) allowing for a simpler reconciliation procedure. In the case where the data are not postselected, both modulation schemes have been proved secure against collective attacks (Ref. \cite{GC06,NGA06,LG10b} for a Gaussian modulation, and Ref. \cite{LG09,LG10} for discrete modulations in the case where the quantum channel is linear). Finally, thanks to a de Finetti theorem in infinite-dimensional Hilbert spaces, it is enough to consider collective attacks to prove the general security of a CVQKD scheme \cite{RC09}.

The typical outline of a CV QKD protocol is the following. Alice prepares $N$ coherent states $|q_k + i p_k\rangle$ where $q_k$ and $p_k$ are real random variables following the appropriate probability distribution: $q_k, p_k$ can be either centered normal random variables or Bernoulli random variables depending on the modulation of the protocol.
Bob measures each state with a homodyne or a heterodyne detection (in the first case, he needs to inform Alice of his choice of quadrature). At this point, Alice and Bob possess $N$ (or $2N$) couples of correlated data $(x_k,y_k)$. These data are related through: $y_k = t x_k + z_k$ where $t$ is an unknown constant and $z_k$ is a centered random variable with unknown variance $\sigma^2$. Alice and Bob then proceed with the parameter estimation procedure whose goal is to estimate both $t$ and $\sigma^2$ by publicly revealing part of their data \cite{LGG10,lev09}. Note that this estimation can never be perfect in practice. The remaining data $(x_k,y_k)$ for $k \in \{1, \cdots, n\}$ are used to distill a secret key. This is done by first applying a reverse reconciliation technique \cite{GG02} where Bob sends some side information to Alice to help her guess the value of $(y_1, \cdots, y_n)$. The side information is typically composed of continuous data, for instance the absolute value $|y_k|$ of Bob's data, as well as of the syndrome of a linear error correcting code (for instance of a Low-density Parity-check (LDPC) code \cite{RSU01,RU02}). The reconciliation procedure is characterized by its efficiency $\beta$ which is the ratio between the length of the bit-string Alice and Bob manage to agree on and the mutual information $I(x;y)$ they initially shared. Finally the privacy amplification allows them to transform this partially unsecure bit-string into a secret key of length $l=n K$ where the asymptotic key rate $K$ is given by:
\begin{equation}
K = \beta I(x;y)-\chi(y;E).
\end{equation}
Taking into account finite size effects leads to a more complicated expression which can be found elsewhere \cite{SR08,LGG10,lev09} (see also discussion below). The quantity $\chi(y;E)$ refers to the Holevo information between the eavesdropper and Bob's data. Note that the main contribution to finite size effects comes from the inaccuracy in the parameter estimation: one should indeed consider for $\chi(y;E)$ the maximal value compatible with the estimation except for some small probability $\epsilon_{PE}$, say $10^{-10}$.

The main limitation in terms of range for CV QKD stems from the finite reconciliation efficiency, especially for a Gaussian modulation in the low signal-to-noise ratio (SNR) regime. Using a discrete quaternary modulation improves the performances significantly as one is now able to perform an efficient reconciliation, even for arbitrarily low SNR \cite{LG09,LG10}. On the other hand, upper bounding the Holevo quantity is more challenging in this scenario, and tight bounds are only available when the modulation variance is small. The reason for it is that the bounds are obtained from an optimality property of Gaussian states \cite{WGC06,GC06}, and that the four-state protocol is close to a Gaussian protocol for low modulation variance only (typically the optimal variance corresponds to sending coherent states with a mean photon number between 0.2 and 0.5).
While this is perfectly fine in theory, it certainly makes the experimental implementation more challenging. 

In this paper, we introduce a new continuous-variable QKD protocol combining an efficient reconciliation procedure and a much tighter bound on $\chi(y;E)$. This protocol outperforms all known practical CV QKD protocols, both in terms of rate and achievable range. It  also allows for larger modulation variances, hence significantly simplifying the experimental implementation for long distances. We will establish the security of this protocol against linear attacks (for instance Gaussian attacks) in the asymptotic regime. By definition, a linear attack corresponds to any action of the eavesdropper compatible with a linear quantum channel between Alice and Bob (see Appendix \ref{linear} for details concerning linear channels). The general case of collective attacks will be treated elsewhere \cite{LG11}. In order to show the robustness of the protocol,  we will also present its performances in a non-asymptotic regime, where the imperfect parameter estimation is taken into account using the techniques described in refs. \cite{LGG10,lev09}.

\section{A new modulation scheme}
Let us first say a few words concerning the reconciliation procedure. A necessary condition in order to achieve long distances is to be able to have an efficient reconciliation at low SNR. The main difficulty here lies in the fact that we need a reverse reconciliation. Indeed, the side information sent by Bob must help Alice without giving Eve any relevant information. 
The only schemes where side information seems to have these properties are the Gaussian modulation where side information describes rotations in $\mathbb{R}^8$ \cite{LAB08} and the binary and quaternary modulations where side information consists of the absolute value of Bob's measurement result \cite{LG09}. 

In order to increase the secret key rate, one needs to find the best possible balance between a large value of $\beta I(x;y)$ and a small value of $\chi(y;E)$. 
From this perspective, the protocol with a Gaussian modulation and the four-state protocol appear to be at the two ends of the spectrum. 
A Gaussian modulation, on one hand, insures the lowest possible value for the upper bound on $\chi(y;E)$, but unfortunately, the quantity $\beta I(x;y)$ is also quite small, and one cannot distill secret keys over large distances with this protocol.
The 4-state protocol, on the other hand, is designed specifically to maximize the quantity $\beta I(x;y)$ at the cost of increasing the provable upper-bound on $\chi(y;E)$, which is a consequence of the fact that a quaternary modulation only roughly approximates a genuine Gaussian modulation for low modulation variances. 

The idea of the protocol presented here is to combine these two solutions to find a better trade-off. 
The modulation scheme now consists in generating points centered on an 7-dimensional sphere in $\mathbbm{R}^8$ (this is done by considering together 4 successive coherent states in phase space). Then, using the same technique as in Ref. \cite{LAB08}, one can reduce the reconciliation problem to the discrete case, which can be efficiently solved as in Ref. \cite{LG09}. However, because the continuous modulation on a sphere in $\mathbb{R}^8$ approximates a Gaussian modulation quite accurately, the bound on $\chi(y;E)$ becomes much tighter than for the four-state protocol.

We now give a detailed description of our new protocol. Alice sends $4N$ coherent states to Bob such that the coordinates of all quadruples $\{|\alpha_{4k}\rangle,  |\alpha_{4k+1}\rangle, |\alpha_{4k+2}\rangle, |\alpha_{4k+3}\rangle\}$ for $k\in \{1, \cdots, N\}$ 
are drawn with the uniform probability on the seven-dimensional sphere of radius $2 \alpha$ in phase space\footnote{This can be done quite simply: Alice only needs to draw eight random variable with a normal probability distribution and then to normalize this eight dimensional vector so that it belongs to the sphere $\mathcal{S}^7$ of radius $2\alpha$ in $\mathbb{R}^8$.}:
\begin{multline}
\label{modulation_octonions}
\mathcal{S}^7 \equiv \{(\alpha_{4k}, \alpha_{4k+1}, \alpha_{4k+2}, \alpha_{4k+3})\in \mathbb{C}^4 \quad \mathrm{such} \quad \mathrm{that}\\
|\alpha_{4k}|^2+|\alpha_{4k+1}|^2+|\alpha_{4k+2}|^2+|\alpha_{4k+3}|^2 = 4 \alpha^2\}.
\end{multline} 
$\alpha$ is related to Alice's modulation variance $V_A$ through $V_A = 2\alpha^2$ (expressed in shot noise units).
Then Bob proceeds with an \emph{heterodyne measurement} (as in Ref. \cite{WLB04} for instance). Here, it is crucial that both quadratures are measured in order to use the property of Eq. \ref{modulation_octonions}. The parameter estimation procedure now consists in revealing $N-n$ quadruples in order to estimate the parameters $t$ and $\sigma^2$ as before. Then, the reconciliation procedure is a mix between the reconciliation using the octonions presented in Ref. \cite{LAB08} and the one described in Ref. \cite{LG09} using the concatenation of good error correcting codes with a repetition code in order to be able to work at very low SNR. It goes at follows. Bob first puts together his $n$ 8-dimensional real vectors ${\bf y^k} = (y_1^k, \cdots, y_8^k)$ and chooses randomly $n$ 8-bit strings $(u_1^k, \cdots, u_8^k)$. These 8-bit strings are mapped on points on a hypercube in $\mathbb{R}^8$ with coordinates ${\bf u^k} = ((-1)^{u_1^k} \frac{{\bf ||{\bf y^k}||}}{2\sqrt{2}},\cdots, (-1)^{u_1^k} \frac{{\bf ||{\bf y^k}||}}{2\sqrt{2}})$ where $||y^k||^2=(y_1^k)^2+\cdots+(y_8^k)^2$.
He then computes the $n$ rotations in $\mathbb{R}^8$ mapping ${\bf y^k}$ to ${\bf u^k}$ as described in Ref. \cite{LAB08} and sends them, together with the value of $||{\bf y^k}||$ to Alice on the authenticated classical channel. Alice applies the same $n$ rotations to her data. At this point, Bob computes the syndrome of his $8n$-bit string for a code $C$ he and Alice agreed on beforehand and sends this syndrome to Alice. This syndrome defines a subset of the $8n$-dimensional hypercube containing the point $({\bf u^1}, \cdots, {\bf u^n})$. If the code $C$ is well chosen, with high probability, Alice recovers the value of $(u_1^k, u_2^k, \cdots, u_n^k)$. The efficiency of this procedure is the same as the one of the reconciliation of 4-state protocol. Alice and Bob can then proceed with privacy amplification to obtain their secret key. 

\section{Performance and security of the protocol}

Our goal here is to evaluate the secret key rate $K$. The first term $\beta I(x;y)$ is rather easy to estimate. Because of the specific reconciliation procedure, $\beta$ is the same as for the discrete-modulation protocol, and can be assumed to be at least $0.8$ for any SNR lower than 1 \cite{LG09}. The mutual information between Alice and Bob corresponds to the capacity of a binary input additive white Gaussian noise channel, which is a function of the SNR.  

\begin{figure}[th]
  \centerline{
    \includegraphics[width=0.45\linewidth]{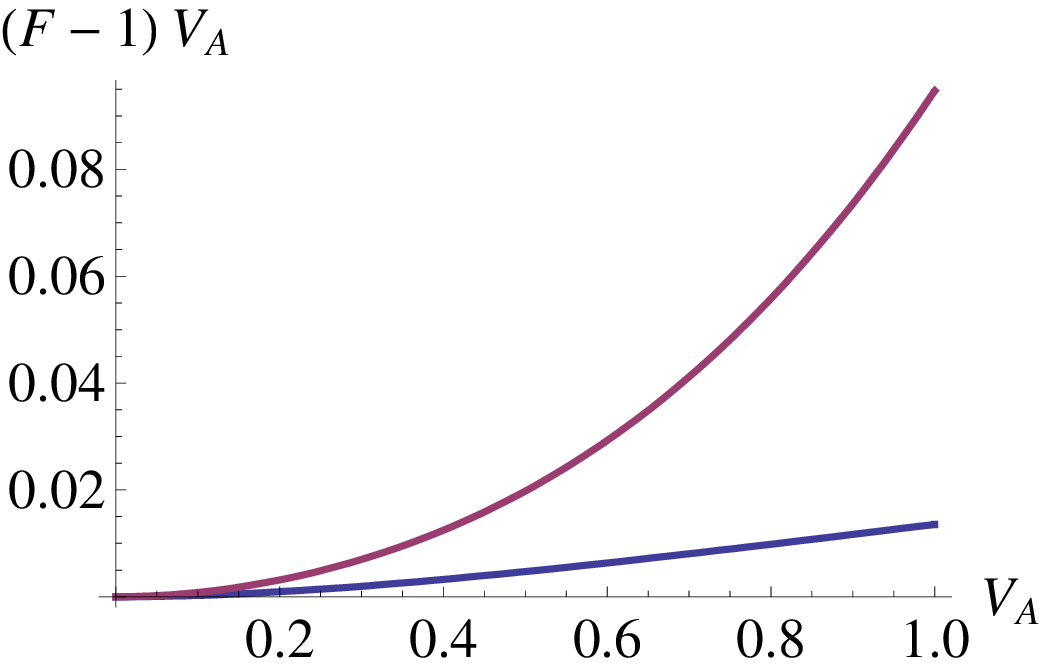}
    \includegraphics[width=0.45\linewidth]{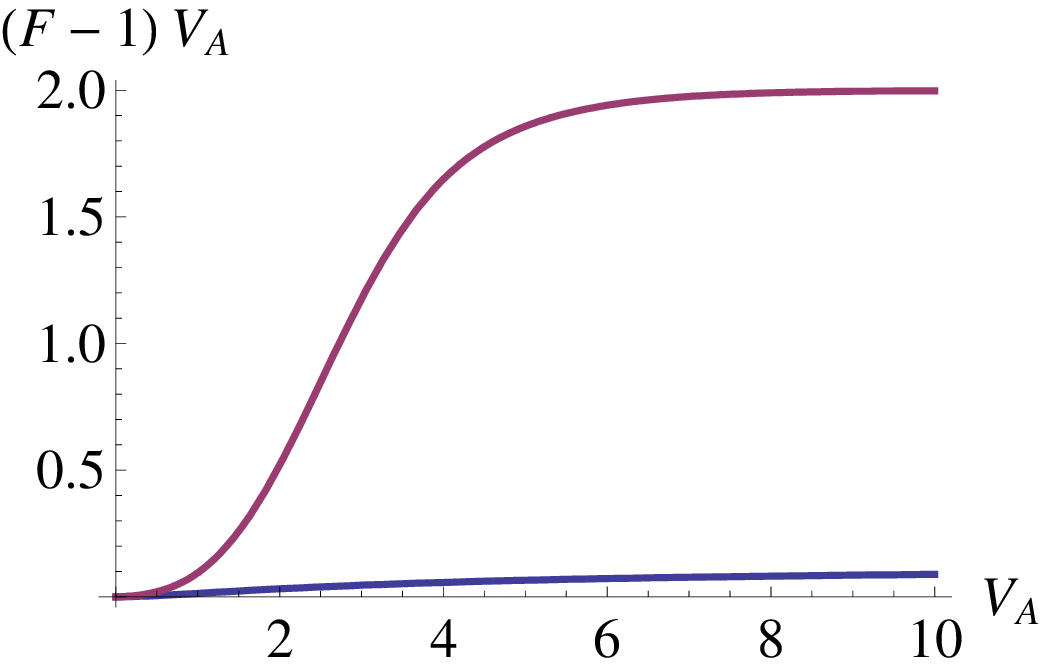}}
\caption{\label{DeltaXi_VA} Equivalent excess noise due to the non-Gaussian modulation. Upper curve refers to the 4-state protocol \cite{LG09}, lower curve to the new continuous-modulation protocol. An excess noise of one unit of shot noise corresponds to an entanglement-breaking channel, therefore no security is possible with such a level of noise.}
\end{figure}

In order to upper bound $\chi(y;E)$, one needs to consider the entanglement-based version of the protocol. Such a ``virtual entanglement'' does not have to be implemented, but it is formally equivalent to the used prepare-and-measure protocol. In this version, Alice starts by preparing $n$ bipartite states 
\begin{equation}
|\Psi\rangle = e^{-2\alpha^2} \sum_{k=0}^{\infty} \frac{(2\alpha)^k}{\sqrt{k!}} \, |\psi_k^4\rangle,
\end{equation}
where
\begin{equation*}
|\psi_k^4\rangle = \frac{1}{\sqrt{{k+3 \choose 3}}} \sum_{\sum_i k_i = k } |k_1,k_2,k_3,k_4\rangle | k_1,k_2,k_3,k_4 \rangle
\end{equation*} 
and performs a POVM on the first half of her state which projects the second half on the coherent states with the right modulation. These coherent states are then sent to Bob. The covariance matrices of this state $|\Psi\rangle$ respectively before and after the transmission through a \emph{linear} channel of transmission $T$ and excess noise $\xi$ are noted $\Gamma^{0} \otimes \mathbbm{1}_4$ and $\Gamma \otimes \mathbbm{1}_4$ with
\begin{equation*}
\Gamma^{0}\! =\! \left(\!
\begin{smallmatrix}
(V_A + 1) \mathbbm{1}_{\! 2} & Z\, \sigma_z \\
Z \,\sigma_z & (V_A+1)\mathbbm{1}_{\! 2} \\ 
\end{smallmatrix}\! \right)\! ,
\Gamma \!= \!\left(\!
\begin{smallmatrix} 
(V_A + 1) \mathbbm{1}_{\! 2} & \sqrt{T} \,Z\, \sigma_z  \\
\sqrt{T} \,Z \, \sigma_z & (1 + T V_A + T \xi)  \mathbbm{1}_{\! 2} \\ 
\end{smallmatrix}\! \right)
\end{equation*}
where $V_A = 2\alpha^2$ is Alice's modulation variance in the Prepare and Measure version of the protocol.
The parameter $Z$ characterizes the level of correlation in phase space between the two halves of the states. The maximal value of $Z$ compatible with quantum mechanics is obtained in the case of a two-mode squeezed state and reads $Z_{\text{TMS}} = \sqrt{V_A^2+2V_A}$. This is therefore the relevant value when considering the QKD protocol with a Gaussian modulation. 
In the case of the continuous-modulation protocol introduced here, one has \cite{lev09}:
\begin{equation}
Z = \frac{1}{2} e^{-2V_A} \sum_{k=0}^{\infty} \frac{\sqrt{k+4}}{k!}\, V_A^{k+\frac{1}{2}}.
\end{equation}
The fact that $Z < Z_{\text{TMS}}$ leads to an increase of the upper bound on $\chi(y;E)$ one can derive from a Gaussian optimality argument. In particular, the value of $\chi(y;E)$ one obtains corresponds to the value one would obtain for a Gaussian modulation protocol with a quantum channel characterized by a transmission $T_G = T/F \approx T$, and  an excess noise $\xi_G = F \xi + (F-1) V_A \approx \xi + (F-1) V_A$, where $F \equiv (Z_{\text{TMS}}/Z)^2$.
Since one has $F \approx 1$ for reasonable values of $V_A$, the main effect of the non-Gaussian modulation is the \emph{equivalent excess noise} $\Delta \xi = (F-1) V_A$. Figure \ref{DeltaXi_VA} displays this equivalent excess noise in the case of the protocol presented here, as well as for the 4-state protocol introduced in \cite{LG09}.
In state-of-the-art implementation, the excess noise is typically less than a few percent of the shot noise. This gives a approximate limit for the value of the equivalent excess noise that is acceptable. In particular, for the 4-state protocol, one needs to work with modulation variances below 0.5 units of shot noise. On the contrary, it becomes possible to work with much higher variances in the case of our new protocol. 

\begin{figure}[th]
  \centerline{
    \includegraphics[width=0.45\linewidth]{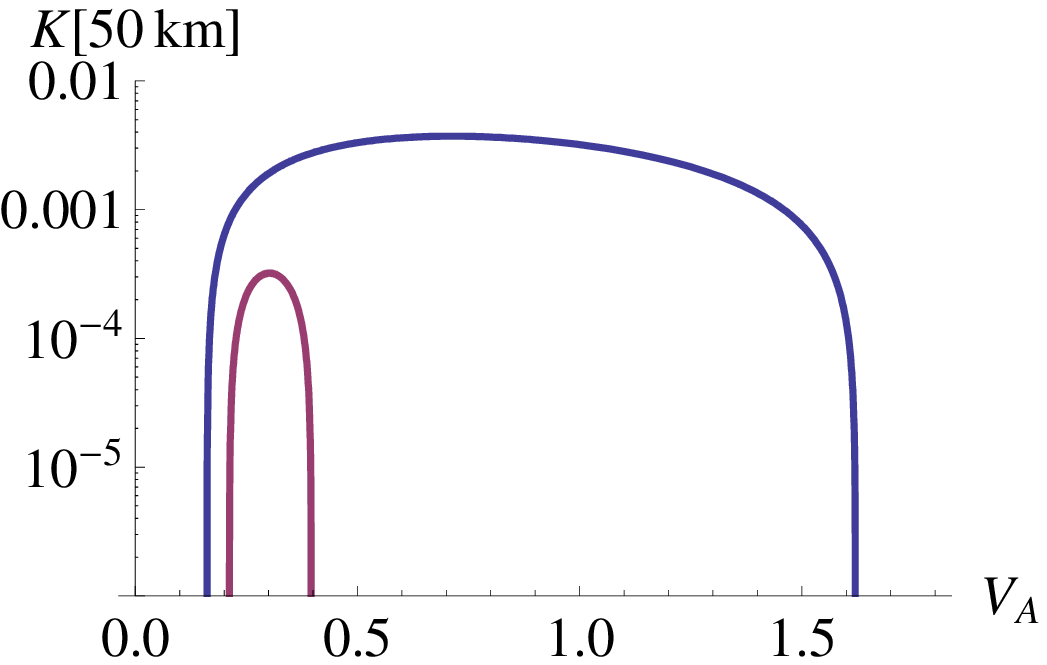}
    \includegraphics[width=0.45\linewidth]{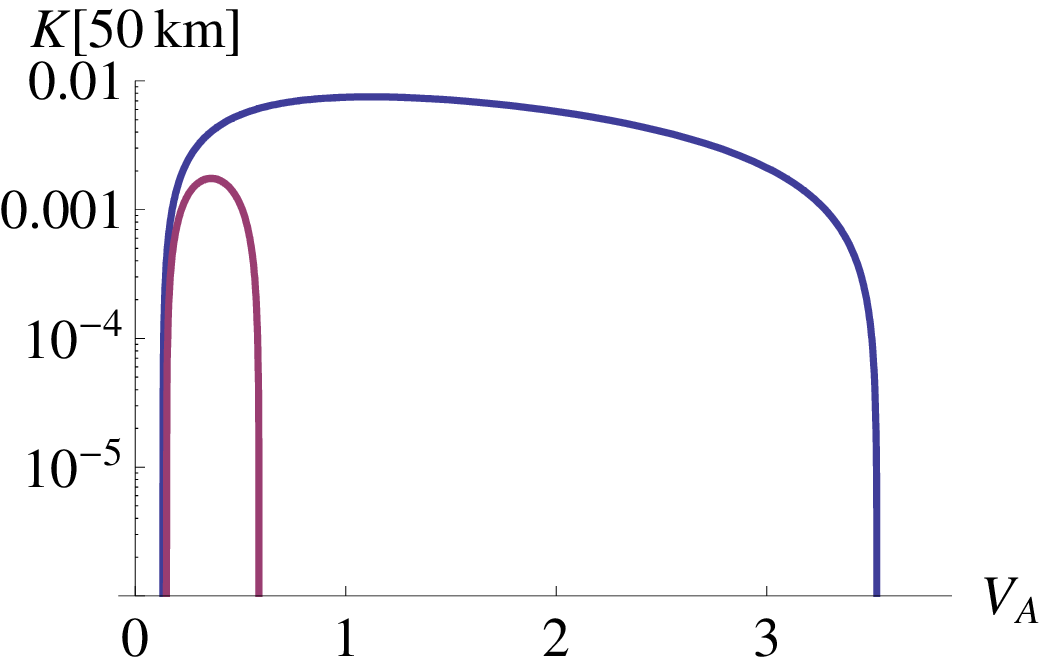}}
 \caption{\label{K4_vs_Kocto_50km_ksi01_eta60} Asymptotic secret key rate for the new protocol and the four-state protocol (heterodyne detection) for a distance of 50 km, as a function of Alice's modulation variance. The various parameters are an excess noise of 0.01 and quantum efficiency of the detectors is $\eta = 60\%$. Reconciliation efficiency is supposed to be a conservative $80\%$ on the left Figure, and an optimistic $90\%$ on the right Figure.}
\end{figure}

This can be seen on Figure \ref{K4_vs_Kocto_50km_ksi01_eta60} where we display the asymptotic secret key rate for a distance of 50 km for the new protocol as well as for the 4-state protocol as a function of Alice's modulation variance. The various parameters are chosen conservatively: a quantum efficiency of $60\%$ and an excess noise of 0.01. Both plots correspond respectively to a reconciliation efficiency of $80\%$ and a more optimistic value of $90\%$. The superiority of the new protocol is quite clear: the secret key rate is higher by nearly an order of magnitude, and one can work with significantly larger modulation variances.  

\begin{figure}[th]
  \centerline{
    \includegraphics[width=0.6\linewidth]{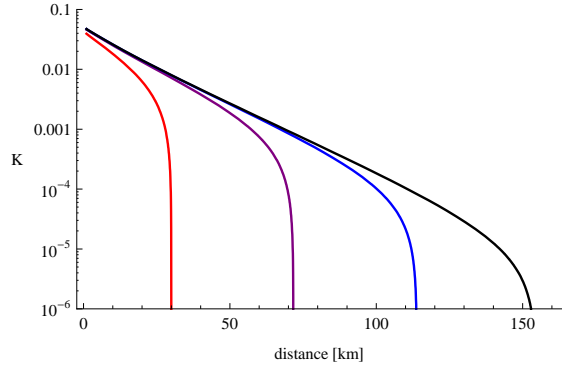}}
\caption{\label{KoctoFinite} Non-asymptotic secret key rate for the new protocol, obtained for realistic  values:  excess noise $\xi = 0.005$, security parameter $\epsilon_{\text{PE}} = 10^{-10}$, quantum efficiency of the detectors  $\eta = 60\%$, reconciliation efficiency $80\%$ for the bi-AWGN channel. Half the samples are used for parameter estimation. From left to right, the block length is equal to $10^{8}, 10^{10}, 10^{12}$ and $10^{14}$.}
\end{figure}

In order to confirm the robustness of the new protocol, we display on  Fig. \ref{KoctoFinite}  the secret key rate when finite size effects are taken into account. The secret key rate is computed against collective attacks, as detailed in Ref. \cite{LGG10}.
Among various finite size effects \cite{SR08}, the most crucial ones for continuous-variable protocols are clearly the imperfect reconciliation efficiency (which prevents the protocol with a Gaussian modulation to achieve key distribution over large distances) and the parameter estimation. While the reconciliation efficiency is taken care of by the 8-dimensional continuous modulation, the parameter estimation is quite sensitive for continuous-variable protocols. In fact, the real problem lies in the estimation of the excess noise $\xi$, which is very small compared to the shot noise, and thus hard to evaluate accurately.

In Fig. \ref{KoctoFinite}, all such finite size effects are taken into account \cite{LGG10}.
The results are rather pessimistic, but remember that this is also true for all discrete-variable protocols \cite{CS09}, and our protocol performs relatively quite well. While exchanging $10^{14}$ quantum signals is rather unrealistic, exchanging $10^8$ or even $10^{10}$ signals can be done with today's technology. Hence, our new protocol allows for the distribution of secret keys over distances of the order of 50 km,  taking into account all finite-size effects.

\section{Perspectives} 

As a conclusion, we presented a new continuous-variable QKD protocol based on a continuous but non-Gaussian modulation and established its security against collective attacks, provided that the quantum channel is linear. The use of a specific reconciliation procedure allows for the distribution of secrets keys over long distances, which was impossible with a Gaussian modulation. Moreover, this protocol clearly outperforms all known practical continuous-variable, with a secret key rate an order of magnitude higher than for the four-state protocol.

An important question at that stage is how to avoid the extra hypothesis that the channel should be linear. As shown in Ref. \cite{LG11}, this can be done by using decoy states, in order to embed the non-Gaussian modulation into an overall gaussian modulation. It is then safe to evaluate the values of $T$ and $\xi$ from a gaussian probe beam, and then to use them as described in the present paper.


\section*{Acknowledgments}
 We acknowledge support from the European Union under project SECOQC (IST-2002-506813) and the ERC Starting grant PERCENT, and from Agence Nationale de la Recherche under projects PROSPIQ (ANR-06-NANO-041-05) and SEQURE (ANR-07-SESU-011-01).

\bibliography{octonion}
\bibliographystyle{unsrt}

\appendix

\section*{Appendix: linear quantum channels}
\label{linear}

We shall define a linear quantum channel by the input-output  relations of the quadrature operators in Heisenberg representation : 
\begin{eqnarray}
X_{out} = g_X X_{in} + B_X  \nonumber \\
P_{out} = g_P  P_{in} + B_P
\end{eqnarray}
where the added noises  $B_X$, $B_P$ are uncorrelated with the input quadratures $X_{in}$, $P_{in} $. 
Such relations have been extensively used for instance in the context of Quantum Non-Demolition (QND) measurements of continuous variables \cite{GLP98}, and they are closely related to the linearized approximation commonly used in quantum optics. Gaussian channels (channels that preserve the Gaussianity of the states) are usual examples of linear quantum channels. However, linear quantum channels may also be non-Gaussian, this will be the case for instance if the added noises $B_X$, $B_P$ are non-Gaussian. 

For our purpose, the main advantage of a linear quantum channel is that it will be characterized by transmission coefficients $T_X = g_X^2$, $T_P = g_P^2$, and by the variances of the added noises $B_X$ and $B_P$. These quantities can be determined even if the modulation used by Alice is non-Gaussian, with the same measured values as when the modulation is Gaussian (because these values are intrinsic properties of the channel). The relevant covariance matrix can then be easily determined, and Eve's information can be bounded by using the Gaussian optimality theorem.

% \begin{thebibliography}{000}

% \bibitem{bib1}

% \bibitem{bib2}

% \bibitem{bib3}

% \end{thebibliography}

\end{document}